\documentclass[a4paper,fleqn]{cas-dc}

\usepackage[numbers]{natbib}
\usepackage{lineno}

\hypersetup{citecolor=black}

\def\tsc#1{\csdef{#1}{\textsc{\lowercase{#1}}\xspace}}
\tsc{WGM}
\tsc{QE}
\tsc{EP}
\tsc{PMS}
\tsc{BEC}
\tsc{DE}

\begin{document}
\let\WriteBookmarks\relax
\def\floatpagepagefraction{1}
\def\textpagefraction{.001}

\title [mode = title]{Observation of temporal routing of topological edge states}                      

\author[1]{Letian Xu}
\credit{Data curation, Formal analysis, Investigation, Software, Validation, Visualization, Writing – original draft}
\author[1]{Muhammad Imran}
\cormark[1]
\ead{imran89123@outlook.com}
\credit{Data curation, Investigation, Visualization, Writing – review \& editing}
\author[1]{Yongtao Jia}
\credit{Resources}
\author[1]{Ying Liu}
\credit{Funding acquisition, Resources}
\author[2]{Huibin Tao}
\cormark[2]
\ead{coldfire2000@mail.xjtu.edu.cn}
\credit{Formal analysis, Investigation, Methodology, Resources}
\author[1]{Rujiang Li}
\cormark[3]
\ead{rujiangli@xidian.edu.cn}
\credit{Conceptualization, Data curation, Formal analysis, Funding acquisition, Investigation, Methodology, Project administration, Resources, Software, Supervision, Validation, Visualization, Writing – review \& editing}

\affiliation[1]{organization={National Key Laboratory of Radar Detection and Sensing, School of Electronic Engineering, Xidian University},
city={Xi'an},
postcode={710071}, 
country={China}}

\affiliation[2]{organization={School of Microelectronics, Xi'an Jiaotong University},
city={Xi'an},
postcode={710049},
country={China}}

\cortext[cor1]{Corresponding author}
\cortext[cor2]{Corresponding author}
\cortext[cor3]{Corresponding author}

\begin{abstract}
The switching or routing of topological edge states is crucial for disorder-immune state transfer. Previous experimental observations have primarily relied on optical waveguide arrays that treat the spatial coordinate as an equivalent temporal dimension. However, direct experimental observation of the routing dynamics of topological edge states in physical time remains challenging.
Here, we report the experimental observation of temporal routing of topological edge states. By constructing a lattice composed of two coupled nonlinear Su-Schrieffer-Heeger (SSH) chains, the topological edge states exhibit temporal routing behavior controlled by nonlinearity. As the excitation voltage increases from the linear regime, we observe that the voltage intensity distribution transitions from periodic interchain exchange to slowed interchain exchange and ultimately to localization with strongly suppressed interchain exchange, thereby validating the phenomenon of temporal routing. 
These dynamics are experimentally observed within a finite time window limited by circuit dissipation.
Our findings provide valuable insights into the temporal manipulation of topological edge states.
\end{abstract}




\maketitle

\section{Introduction}

Topological insulators behave as conventional insulators in the bulk but exhibit conductive boundaries due to the presence of topologically protected edge states \cite{RMP82-3045,RMP83-1057}. Although this concept was originally proposed in condensed matter physics, analogues of topological insulators and a wide range of topological phases have been realized across diverse physical platforms, including acoustics \cite{NRP1-281,NRM7-974,RPP86-106501}, ultracold atomic systems \cite{RMP91-015005}, photonics \cite{RMP91-015006,LSA9-1,APR2-2100013}, and circuit lattices \cite{PR1093-1,APLED,PRL128-097701,RPP89-048004,FP20-044204}. The immunity of topological edge states to local deformations and disorder is promising for the design of various types of devices \cite{NRP4-184,NC10-2058,AM30-1803229,sa9-eadi,LPR19-2400567,NE5-300,SA10-eadp,science358-636,science359-1231,nature578-246,nphys20-852,PQW1-8}. Among these applications, the switching or routing of topological edge states is particularly significant because it enables state transfer in a controllable and disorder-immune manner. To date, various schemes for the switching or routing of topological edge states have been proposed. Typical examples include the adiabatic pumping of topological edge states \cite{PRL109-106402,LPR10-995,nature553-59,PNAS121}, Bloch oscillations induced by a gradient force \cite{PRA99-053814}, edge-state switching achieved through temporal periodic modulation of the lattice depth \cite{LPR12-1700348,OL44-3342}, coupling of topological edge states in small-size lattices \cite{PRA104-053531,LPR14-1900193,OL49-6377}, a topological coupler composed of two mirror-symmetric staggered helical lattices \cite{OL46-6067}, nonlinearity-controlled switching of topological edge states \cite{nano11-3653}, {and loss-controlled switching between topological edge states via the macroscopic Zeno effect \cite{LPR17-2300024}}.

Typically, the experimental observation of the switching or routing of topological edge states relies on optical waveguide arrays \cite{PRL109-106402,nature553-59,PNAS121,LPR14-1900193,OL49-6377,nano11-3653,LPR17-2300024}. By treating the longitudinal propagation coordinate of the waveguide arrays as an equivalent temporal dimension, the temporal evolution of a state can be simulated through the spatial light evolution in the waveguide array. This analogy has facilitated the design of experimentally realistic structures based on tight-binding models and has addressed the experimental challenge of directly observing the fast temporal evolution of states. However, despite these advantages, the dynamic light evolution in a waveguide array is often not visually observable, with only the output field available for experimental measurement \cite{LSA13-99}. Moreover, it is important to note that
{the evolution in these waveguide arrays is encoded in the longitudinal propagation coordinate rather than physical time.
Therefore, direct experimental observation of the routing dynamics of topological edge states in physical time remains challenging.}

In this paper, we experimentally demonstrate the temporal routing of topological edge states in a circuit lattice. We construct a nonlinear circuit lattice composed of two coupled Su-Schrieffer-Heeger (SSH) chains with onsite nonlinearity and excite the rightmost site of the left SSH circuit chain at the initial moment. The voltage intensity distributions across the entire circuit lattice reveal the nonlinearity-controlled temporal routing of topological edge states. Specifically, in the linear regime, the voltage intensity exhibits periodic exchange between the left and right chains. As the excitation voltage increases, the enhanced nonlinearity results in slowed interchain exchange. At sufficiently high excitation voltages, the interchain exchange is progressively suppressed, leading to persistent localization at the excitation site. Consequently, the transition from periodic interchain exchange to slowed interchain exchange, and ultimately to localization with strongly suppressed interchain exchange, validates our observation of nonlinearity-controlled temporal routing of topological edge states. 
{Our circuit platform enables direct observation of the routing dynamics of topological edge states in physical time, whereas the longitudinal propagation coordinate serves as the evolution variable in optical waveguide arrays \cite{nano11-3653}.}
Our findings provide valuable insights into the temporal manipulation of topological edge states and may have applications in the design of devices such as topological couplers and routers.

\section{Model}

\begin{figure}
\centering
\includegraphics[width=9cm]{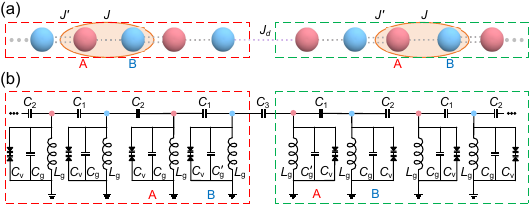}
\caption{Tight-binding lattice model and corresponding circuit lattice.
(a) A nonlinear lattice composed of two identical Su-Schrieffer-Heeger (SSH) chains with onsite nonlinearity. The left SSH chain (outlined by the red rectangle) and the right SSH chain (outlined by the green rectangle) are coupled through the interchain hopping $J_d$. Within each SSH chain, the intracell and intercell hoppings are denoted as $J$ and $J^{\prime}$, respectively. The onsite energies at the sublattice sites $\mathrm{A}$ and $\mathrm{B}$ are wavefunction-dependent.  
(b) Circuit lattice corresponding to the tight-binding lattice model. Each $LC$ oscillator, consisting of an inductor $L_\mathrm{g}$, a capacitor $C_\mathrm{g}$ (or $C_\mathrm{g}^{\prime}$), and a common-cathode varactor diode $C_\mathrm{v}$, emulates a lattice site as shown in (a). Within the left or right circuit chain, the couplings provided by capacitors $C_{1}$ and $C_{2}$ correspond to the intracell and intercell hoppings in the SSH chain, respectively. The interchain coupling contributed by capacitor $C_{3}$ plays the role of interchain hopping between the SSH chains.}
\label{fig1}
\end{figure}

We consider a nonlinear lattice composed of two identical SSH chains \cite{PRL42-1698,OL51-1649}. As illustrated in Fig.~\ref{fig1}(a), the left SSH chain (outlined by the red rectangle) and the right SSH chain (outlined by the green rectangle) are coupled through the interchain hopping $J_d$. Within each SSH chain, the intracell and intercell hoppings are denoted as $J$ and $J^{\prime}$, respectively. The onsite energies of both the sublattice sites $\mathrm{A}$ and $\mathrm{B}$ are nonlinear \cite{PRR8-L012,PRA113-053506}. Specifically, the onsite energy at any given lattice site depends on the wavefunction at that site.

The lattice model illustrated in Fig. \ref{fig1}(a) can be physically realized using a circuit lattice. As depicted in Fig. \ref{fig1}(b), each $LC$ oscillator, consisting of an inductor $L_\mathrm{g}$, a capacitor $C_\mathrm{g}$ (or $C_\mathrm{g}^{\prime}$), and a common-cathode varactor diode $C_\mathrm{v}$, emulates a lattice site as shown in Fig. \ref{fig1}(a). In the left or right circuit chain, the couplings provided by capacitors $C_{1}$ and $C_{2}$ correspond to the intracell and intercell hoppings in the SSH chain, respectively. The interchain coupling from capacitor $C_{3}$ represents the interchain hopping between SSH chains. Therefore, the circuit lattice, as shown in Fig. \ref{fig1}(b), serves as a direct realization of the tight-binding lattice model presented in Fig. \ref{fig1}(a). The voltage envelopes at the circuit nodes (denoted as red and blue dots) are analogous to the wavefunctions at the lattice sites, and they satisfy the following equations (see derivations in Appendix \ref{app_model}):
\begin{eqnarray}
\mathrm{i}\frac{\mathrm{d}}{\mathrm{d}t} V_{n}^{\sigma,\mathrm{A}} &=& E\left(V_{n}^{\sigma,\mathrm{A}}\right)V_{n}^{\sigma,\mathrm{A}} + J V_{n}^{\sigma,\mathrm{B}} + J^{\prime} V_{n-1}^{\sigma,\mathrm{B}} \nonumber \\
&&+ \delta_{\sigma,\mathrm{R}}\delta_{n,1}J_{d}V_{N}^{\mathrm{L},\mathrm{B}},  \label{eq1}\\
\mathrm{i}\frac{\mathrm{d}}{\mathrm{d}t} V_{n}^{\sigma,\mathrm{B}} &=& E\left(V_{n}^{\sigma,\mathrm{B}}\right)V_{n}^{\sigma,\mathrm{B}} + J V_{n}^{\sigma,\mathrm{A}} + J^{\prime} V_{n+1}^{\sigma,\mathrm{A}} \nonumber \\
&&+ \delta_{\sigma,\mathrm{L}}\delta_{n,N}J_{d}V_{1}^{\mathrm{R},\mathrm{A}}. \label{eq2}
\end{eqnarray}
Note that, to ensure that the onsite energies at all lattice sites are equal  in the linear limit, we set $C_{g}^{\prime}=C_{g}+C_{2}-C_{3}$. 

In Eqs. (\ref{eq1})-(\ref{eq2}), $E\left(V_{n}^{\sigma,\tau}\right) = E_{0}+g\left(V_{n}^{\sigma,\tau}\right)$ ($\tau = \mathrm{A}, \mathrm{B}$) is the onsite energy, which includes the constant term $E_0 = \left[1- \left(C_1+C_2 \right) / \left(2 C_\Sigma \right) \right] \omega_{0}$ and the voltage-dependent term $g\left(V_{n}^{\sigma,\tau}\right) = -C_{\mathrm{NL}} \left(V_{n}^{\sigma,\tau}\right) /\left(2 C_\Sigma \right) \omega_0$. Similarly, $J = C_{1}/\left(2 C_\Sigma \right) \omega_0$ and $J^{\prime} = C_{2}/\left(2 C_\Sigma \right) \omega_0$ represent the intracell and intercell coupling coefficients within each SSH circuit chain. Additionally, $J_d = C_{3}/\left(2 C_\Sigma \right) \omega_0$ denotes the interchain coupling coefficient. The parameters in the above expressions are defined as $\omega_{0} = 1/\sqrt{L_\mathrm{g} C_\Sigma}$, $C_{\Sigma} =C_\mathrm{g}+C_{\mathrm{L}}$, and $C_{\mathrm{v}}\left(V_{n}^{\sigma,\tau}\right)=C_{\mathrm{L}}+C_{\mathrm{NL}} \left(V_{n}^{\sigma,\tau}\right)=C_\mathrm{L}/\left(1+\left\vert V_{n}^{\sigma,\tau} /v_{0} \right\vert \right)^{M}$, where $v_{0}$ and $M$ are constants determined by the specific diode model. 
{
This form of the onsite nonlinearity follows from the experimentally measured capacitance-voltage relation of the common-cathode diodes (see Appendix \ref{app_exp}).}
Furthermore, in Eqs. (\ref{eq1})-(\ref{eq2}), $n=1, \cdots , N$ denotes the cell index within the left or right chain, with $N$ representing the number of unit cells in each circuit chain. The symbol $\sigma$ indicates the chain index, which can be either $\mathrm{L}$ (for the left chain) or $\mathrm{R}$ (for the right chain). Lastly, $\delta_{\sigma,\mathrm{L}}$, $\delta_{\sigma,\mathrm{R}}$, $\delta_{n,1}$, and $\delta_{n,N}$ are the Kronecker delta functions.

\section{Results}

\subsection{Temporal routing of topological edge states}
\label{sec_temporal}

For the lattice model depicted in Fig. \ref{fig1}(a), in the linear limit and when $J_{d}=0$, if the intracell hopping $J$ is smaller than the intercell hopping $J^{\prime}$, the uncoupled SSH chains are both topologically nontrivial and support the topological edge states located at the chain edges \cite{book}. 
{
When $J_{d} \neq 0$, the topological edge state at the right end of the left chain and the topological edge state at the left end of the right chain hybridize into a pair of localized interface modes with symmetric and antisymmetric spatial distributions, respectively \cite{OL51-1649}. We refer to these modes as coupled topological edge states to indicate their origin in the topological edge states of the uncoupled SSH chains.}

\begin{figure}
\centering
\includegraphics[width=9cm]{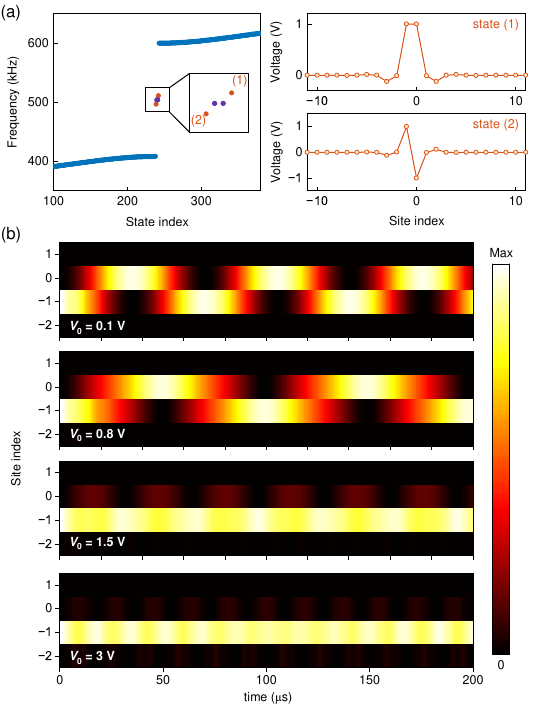}
\caption{{Hybridization of topological edge states and nonlinearity-controlled temporal routing.}
(a) Eigenfrequency spectrum of the circuit lattice in the linear regime. In addition to the bulk states (denoted by blue dots), the two topological edge states localized at the adjacent ends of the uncoupled SSH chains are now hybridized to form two coupled topological edge states (represented by red dots) with symmetric and antisymmetric distributions, respectively. The two purple dots correspond to the edge states located at the ends of the entire circuit lattice. The spatial distributions of the coupled topological edge states are displayed in the insets. (b) Temporal evolution of the voltage intensity distributions when the rightmost site of the left circuit chain is excited by a voltage $V_{0}$. In (a)-(b), negative site indices correspond to the left circuit chain, while zero and positive site indices correspond to the right circuit chain.}
\label{fig2}
\end{figure}

{
Correspondingly, for the circuit lattice shown in Fig.~\ref{fig1}(b), in the linear limit $V_n^{\sigma,\tau}\to 0$ ($\tau=\mathrm{A},\mathrm{B}$), which is equivalent to setting $C_{\mathrm{NL}}=0$, the uncoupled SSH circuit chains are topologically nontrivial when $C_1<C_2$.} We set the values of the circuit components as follows: $L_{\mathrm{g}} = 15~\mu\mathrm{H}$, $C_g = 0$, $C_1 = 180~\mathrm{pF}$, $C_2 = 1500~\mathrm{pF}$, $C_3 = 100~\mathrm{pF}$, and $N = 120$. Meanwhile, the parameters for the common-cathode diodes are $C_{\mathrm{L}}=4.31~\mathrm{nF}$, 
{$v_{0}=1.7~\mathrm{V}$,} and $M=0.31$, as derived from a realistic diode model (see Appendix \ref{app_exp}). Under these circuit parameters, as shown in Fig. \ref{fig2}(a), the states labeled as (1) and (2) (denoted by red dots) in the eigenfrequency spectrum reside within the bandgap created by the bulk states (denoted by blue dots), exhibiting symmetric and antisymmetric voltage distributions, respectively (see the insets of Fig. \ref{fig2}(a)). Specifically, at the two circuit nodes directly connected by capacitor $C_{3}$, the voltage distributions are in-phase for state (1) and out-of-phase for state (2). 
{
The interchain coupling $J_d$ shifts the eigenfrequencies of these two modes away from the midgap frequency. As $J_d$ is increased, these hybridized modes can merge with the bulk bands without closing the bulk gap.}
Note that negative site indices correspond to the left circuit chain, while zero and positive site indices correspond to the right circuit chain. Additionally, it is worth mentioning that the two mid-gap states, denoted by the purple dots in Fig. \ref{fig2}(a), correspond to the edge states located at the ends of the entire circuit lattice, which are not relevant to our discussions here.

With the inclusion of nonlinear effects, the existence of symmetric and antisymmetric coupled topological edge states has previously been utilized to design a nonlinearity-controlled switch based on optical waveguide arrays \cite{nano11-3653}. Since the propagation coordinate of the waveguide arrays is treated as an equivalent temporal dimension, the spatial switching behavior is transformed into the phenomenon of nonlinearity-controlled temporal routing in our circuit lattice.
{
Here, temporal routing refers to the time-dependent redistribution of voltage intensity between the adjacent ends of the two SSH chains through the hybridization of their topological edge states.}

At the initial moment, we apply a voltage $V_{0}$ to the rightmost site of the left SSH circuit chain, i.e., $V_{N}^{\mathrm{L},\mathrm{B}} \left(t=0\right)=V_{0}$. 
{
Here, $V_0$ denotes the tunable initial excitation voltage, while all diode parameters remain fixed.}
When $V_{0}$ is small, this initial excitation can be approximated as a superposition of the symmetric and antisymmetric coupled topological edge states, expressed as $\mathbf{V} \left(t=0\right) \approx V_{0}\left(\boldsymbol{\phi}_{\mathrm{s}}+\boldsymbol{\phi}_{\mathrm{as}}\right)/\sqrt{2}$, where $\mathbf{V} \left(t=0\right)$ is the initial voltage distribution in the entire circuit lattice, and $\boldsymbol{\phi}_{\mathrm{s}}$ and $\boldsymbol{\phi}_{\mathrm{as}}$ represent the voltage distributions of the symmetric and antisymmetric coupled topological edge states, respectively. As time evolves, the voltage distribution in the circuit lattice can be approximated by $\mathbf{V}\left(t\right) \approx V_{0}\left(\boldsymbol{\phi}_{\mathrm{s}}e^{-\mathrm{i}\omega_\mathrm{s}t}+\boldsymbol{\phi}_{\mathrm{as}} e^{-\mathrm{i}\omega_\mathrm{as}t} \right)/\sqrt{2}$, where $\omega_{\mathrm{s}}/2\pi$ and $\omega_{\mathrm{as}}/2\pi$ are the eigenfrequencies of the symmetric and antisymmetric coupled topological edge states, as shown in Fig. \ref{fig2}(a). The voltage intensity at each lattice site oscillates with a frequency of $\left(\omega_{\mathrm{s}}-\omega_{\mathrm{as}}\right)/2\pi$. The first row of Fig. \ref{fig2}(b) displays the voltage intensity distribution obtained from a direct numerical integration of Eqs. (\ref{eq1})-(\ref{eq2}), with the initial excitation set to $V_{0}=0.1~\mathrm{V}$ (see Appendix \ref{app_model} 
for the numerical method). As in the insets of Fig. \ref{fig2}(a), negative site indices correspond to the left circuit chain, while zero and positive site indices correspond to the right circuit chain. The temporal oscillation of voltage intensity shown in the first row of Fig. \ref{fig2}(b) is in agreement with the preceding physical expectations.

As the excitation voltage $V_{0}$ increases, the nonlinear effects in the circuit lattice become more pronounced, rendering the approximations used for $V_{0}=0.1~\mathrm{V}$ no longer valid. As illustrated in the second row of Fig. \ref{fig2}(b) for $V_{0}=0.8~\mathrm{V}$, the increased nonlinearity leads to an elongation of the oscillation period, indicating that the voltage intensity is exchanged between the left and right SSH circuit chains at a lower frequency. This phenomenon aligns with the prior observation of slowed switching speed resulting from increased nonlinearity in waveguide arrays \cite{nano11-3653}.

As the excitation voltage $V_{0}$ continues to increase, the exchange of voltage intensity between the two SSH chains becomes progressively suppressed, ultimately leading to persistent localization at the excitation site. As shown in the third row of Fig. \ref{fig2}(b), for $V_{0} = 1.5~\mathrm{V}$, the voltage intensity is primarily localized at the rightmost site of the left SSH circuit chain. This nonlinearity-induced localization persists even at higher excitation voltages. As illustrated in the fourth row of Fig. \ref{fig2}(b) for $V_{0} = 3~\mathrm{V}$, the exchange of voltage intensity between the two SSH chains is strongly suppressed, and the voltage intensity remains well localized at the excitation site.
 
Therefore, as shown in Fig. \ref{fig2}(b), with an increase in the excitation voltage, the voltage intensity in the circuit lattice transitions from periodic interchain exchange to slowed interchain exchange and ultimately to localization with strongly suppressed interchain exchange. 
{This phenomenon corresponds to nonlinearity-controlled temporal routing of topological edge states in physical time, whereas the spatial switching in optical waveguide arrays is described using the longitudinal propagation coordinate \cite{nano11-3653}.}

{
To quantify the contribution of bulk states to the routing dynamics, we project the voltage vector $\mathbf{V}(t)$ onto the orthonormal linear eigenmodes of the entire lattice~\cite{PRA104-053531}. The maximum total bulk-mode occupation is approximately $0.08$ within the investigated voltage and time ranges, indicating limited participation of the bulk states. Given this limited contribution, we further analyze the occupations of the left- and right-localized states constructed from the symmetric and antisymmetric coupled topological edge states to examine how nonlinearity modifies the interchain exchange. Physically, weak nonlinear frequency shifts at low excitation voltages allow nearly resonant exchange between the two localized states. At high excitation voltages, the unequal amplitudes on the two sides produce different nonlinear frequency shifts, resulting in a frequency mismatch that suppresses interchain exchange and ultimately leads to localization predominantly on the initially excited side (see Appendix~\ref{app_mechanism}).}

{
To further examine the role of the topological edge states, we interchange the intracell and intercell coupling capacitances by setting $C_1=1500~\mathrm{pF}$ and $C_2=180~\mathrm{pF}$, so that both SSH chains are in the topologically trivial phase. At $V_0=0.1~\mathrm{V}$, the initial excitation spreads into the bulk, and the periodic interchain exchange of the voltage intensity between the two interface sites is absent (see Fig.~\ref{app_trivial_fig} in Appendix \ref{app_trivial} for the corresponding temporal evolution). At $V_0=0.8~\mathrm{V}$, the spatial spreading is reduced, but the excitation still extends into the bulk of the left chain. At $V_0=1.5~\mathrm{V}$ and $3~\mathrm{V}$, the voltage intensity is predominantly confined to the excitation site and its neighboring site in the left chain, with oscillations between these two sites. These behaviors at different excitation voltages therefore highlight the role of the topological edge states in supporting the temporal routing observed in Fig.~\ref{fig2}.}

Additionally, we note that in Eqs. (\ref{eq1})–(\ref{eq2}), we do not consider the circuit dissipation in the circuit lattice. However, even if dissipation terms are included in Eqs. (\ref{eq1})–(\ref{eq2}), the temporal routing phenomenon discussed above is still expected to occur (see the model with circuit dissipation in Appendix \ref{app_dissipation}).
 
\subsection{Experimental observation}

\begin{figure}
\centering
\includegraphics[width=9cm]{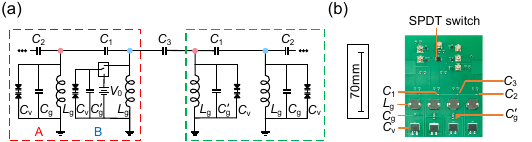}
\caption{Experimental circuit schematic and fabricated circuit sample.
(a) The experimental circuit schematic features a single-pole double-throw (SPDT) switch and a direct current (DC) voltage source connected to the rightmost node of the left circuit chain.
The left and right circuit chains are outlined by red and green rectangles, respectively.
(b) A zoomed-in view of the fabricated printed circuit board (PCB) for the circuit lattice is shown, with the key circuit components labeled. A scale bar for the PCB is included.}
\label{fig3}
\end{figure}

To experimentally observe the temporal routing of topological edge states, we introduce a single-pole double-throw (SPDT) switch and a direct current (DC) voltage source connected to the rightmost node of the left circuit chain. As shown in Fig.~\ref{fig3}(a), when the capacitor \( C_{g}^{\prime} \) and varactor diode \( C_{\mathrm{v}} \) are connected to the DC voltage source \( V_{0} \), they will charge to the voltage \( V_{0} \). This charging process corresponds to the preparation of the initial state $\mathbf{V} \left(t=0\right)$. Once the charged capacitor and diode are switched to the inductor \( L_{\mathrm{g}} \), they will discharge into the entire circuit lattice, which corresponds to the temporal evolution. This two-step process, commonly referred to as quench dynamics in circuit lattices, has been previously implemented to observe edge solitons in a trimer lattice \cite{CP8} and various localized states in a nonlinear quadrupole topological insulator \cite{CSF207}. Figure \ref{fig3}(b) presents the fabricated printed circuit board (PCB) of the circuit lattice, with the key circuit components labeled in the inset (see Appendix \ref{app_exp} for experimental details).

\begin{figure}
\centering
\includegraphics[width=9cm]{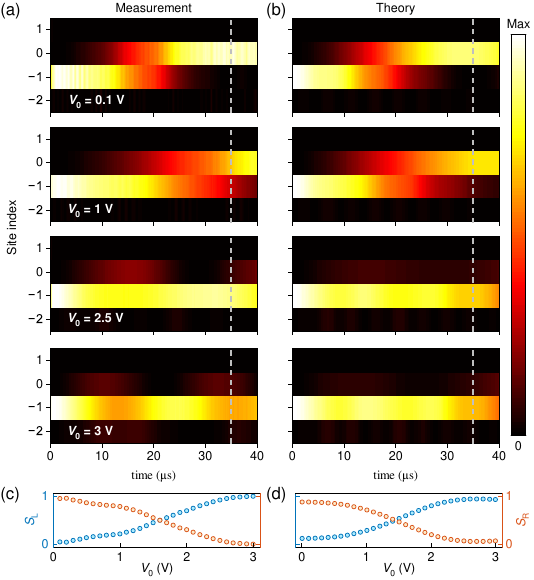}
\caption{Temporal evolution of voltage intensities under different excitation voltages and corresponding power ratios at a fixed time of $t = 35~\mu\mathrm{s}$.
(a) Experimental measurements of the temporal evolutions of voltage intensities under varying excitation voltages.  
(b) Theoretical results corresponding to the same parameters.  
(c) Experimental and (d) theoretical power ratios $S_\mathrm{L}$ and $S_\mathrm{R}$ evaluated at $t = 35~\mu \mathrm{s}$ plotted against the excitation voltage $V_{0}$. For reference, the time $t = 35~\mu \mathrm{s}$ is marked by gray dashed lines in (a) and (b).
}
\label{fig4}
\end{figure}

The experimentally measured temporal evolution of voltage distributions for different excitation voltages is shown in Fig.~\ref{fig4}(a). For comparison, the theoretical results are presented in Fig.~\ref{fig4}(b). The results indicate that when $V_{0} = 0.1~\mathrm{V}$, we observe clear periodic exchange of the voltage distributions between the left and right circuit chains. The left chain corresponds to negative site indices, while the right chain corresponds to zero and positive site indices. As the excitation voltage increases to $V_{0} = 1~\mathrm{V}$, the oscillation period enlarges due to enhanced nonlinearity, implying that the voltage intensity is exchanged between the left and right SSH circuit chains at a lower frequency. This observation validates the transition from periodic interchain exchange to slowed interchain exchange. It is worth noting that the rapid temporal oscillations of the voltage distributions are due to experimental voltage noise (see more explanations in Appendix \ref{app_exp}). Besides, the theoretical results are obtained from a model with circuit dissipation (see details in Appendix \ref{app_dissipation}).

When the excitation voltage is further increased to $V_{0} = 2.5~\mathrm{V}$, the interchain exchange of voltage intensity is progressively suppressed, ultimately leading to persistent localization at the rightmost node of the left SSH circuit chain, as shown in Figs.~\ref{fig4}(a) and (b). At $V_{0} = 3~\mathrm{V}$, the interchain exchange is strongly suppressed, resulting in the voltage intensities being strongly localized at the excitation site. 
{
Thus, as the excitation voltage increases, a transition is experimentally observed from periodic interchain exchange to slowed interchain exchange, and ultimately to localization with strongly suppressed interchain exchange, although circuit dissipation shifts the observed transition to a higher excitation voltage.}
We refer to this behavior as nonlinearity-controlled temporal routing of topological edge states, which has been validated by the experimental observations.

To quantitatively describe the temporal routing of topological edge states, we define the ratio of the power in a single SSH circuit chain to the total power in the circuit lattice as follows:
\begin{equation}
S_{\sigma} (t) = \frac{\sum_{n,\tau} |V_{n}^{\sigma,\tau}(t)|^{2}}{\sum_{n, \sigma, \tau} |V_{n}^{\sigma,\tau}(t)|^{2}},
\end{equation}
where $n=1, \ldots , N$, $\sigma= \mathrm{L}, \mathrm{R}$, and $\tau = \mathrm{A}, \mathrm{B}$. We evaluate the power ratios $S_{\mathrm{L}}$ and $S_{\mathrm{R}}$ at $t = 35~\mu \mathrm{s}$, which corresponds to the moment when the voltage intensity at the excitation site reaches its minimum for $V_{0} = 0.1~\mathrm{V}$. From the experimental measurements shown in Fig.~\ref{fig4}(c), $S_{\mathrm{L}}$ attains its minimum value at $V_{0} = 0.1~\mathrm{V}$, while $S_{\mathrm{R}}$ reaches its maximum. This indicates that the remaining power is predominantly concentrated in the right SSH chain at this time. As the excitation voltage $V_{0}$ increases, $S_{\mathrm{L}}$ rises while $S_{\mathrm{R}}$ declines. This observation is consistent with the slowed interchain exchange depicted in Fig.~\ref{fig4}(a). When the excitation voltage is further increased to approximately $V_{0} = 2.5~\mathrm{V}$, $S_{\mathrm{L}} \approx 1$ and $S_{\mathrm{R}} \approx 0$, implying that the power is strongly localized at the left SSH circuit chain. The strong localization in the left SSH circuit chain is preserved as the voltage continues to increase. For comparison with the experimental measurement results, Fig.~\ref{fig4}(d) presents the theoretical results derived from the voltage intensities at the same moment. The agreement between the experimental results in Fig.~\ref{fig4}(c) and the theoretical results in Fig.~\ref{fig4}(d) further validates the temporal routing of topological edge states.

\section{Conclusion}

In conclusion, we have experimentally observed the temporal routing of topological edge states in a nonlinear circuit lattice composed of two coupled SSH chains with onsite nonlinearity. By exciting the end site of one SSH circuit chain, we have demonstrated that the voltage intensity transitions from periodic interchain exchange to slowed interchain exchange, ultimately reaching persistent localization at the excitation site as nonlinearity increases. Specifically, in the linear regime, the voltage intensity exhibits periodic exchanges between the left and right chains. As the excitation voltage increases, the enhanced nonlinearity leads to slowed interchain exchange. At sufficiently high excitation voltages, these interchain exchanges are progressively suppressed, resulting in persistent localization at the excitation site. 
{In our circuit lattice, the routing dynamics of topological edge states are directly observed in physical time, whereas the longitudinal propagation coordinate serves as the evolution variable in optical waveguide arrays \cite{nano11-3653}.}
Our findings provide valuable insights into the temporal manipulation of topological edge states and may have implications for the design of devices such as topological couplers and routers.

\appendix

\section{Circuit equations}
\label{app_model}

For common-cathode varactor diodes, under a monochromatic voltage $U(t)=\frac{1}{2}v e^{-\mathrm{i}\omega t}+\mathrm{c.c.}$, the capacitance of the diode can be expressed as $C_{\mathrm{v}}(v)=C_{\mathrm{L}}+C_{\mathrm{NL}}(v)=C_{\mathrm{L}}/\left(1+\left\vert v/v_{0}\right\vert\right)^{M}$ \cite{PRB113-214310}. Here, $v_{0}$ and $M$ are constants determined by the specific diode model.

Then, we derive the circuit equations for the lattice shown in Fig.~\ref{fig1}(b). By applying Kirchhoff's current law to each circuit node in the left or right SSH circuit chain, 
we obtain a set of second-order differential equations \cite{PRB113-214310}. For instance, for the sublattice site $\mathrm{A}$ in the $n$th unit cell of the left SSH circuit chain with $n\neq N$, we have
\begin{eqnarray}
&&L_\mathrm{g} \!\left[ C_{\Sigma} + C_1+C_2+C_{\mathrm{NL}}\!\left(v_{n}^{\mathrm{L},\mathrm{A}}\right)\right]\!\frac{\mathrm{d}^2 U_{n}^{\mathrm{L},\mathrm{A}}}{\mathrm{d}t^2} \nonumber \\
&-&L_\mathrm{g}C_1\frac{\mathrm{d}^2U_{n}^{\mathrm{L},\mathrm{B}}}{\mathrm{d}t^2} 
-L_\mathrm{g}C_2\frac{\mathrm{d}^2 U_{n-1}^{\mathrm{L},\mathrm{B}}}{\mathrm{d}t^2}+U_{n}^{\mathrm{L},\mathrm{A}}=0,
\label{eq_app_A}
\end{eqnarray}
where $C_\Sigma=C_\mathrm{g}+C_\mathrm{L}$ is defined. When the condition $C_{\Sigma}\gg C_{1},\,C_{2},\,C_{3},\,C_{\mathrm{NL}}$ holds, the voltage is quasi-monochromatic and can be expressed as
$U_{n}^{\mathrm{L},\mathrm{A}}(t)=\frac{1}{2}v_{n}^{\mathrm{L},\mathrm{A}}(t)e^{-\mathrm{i}\omega_{0}t}+\mathrm{c.c.}$,
where $v_{n}^{\sigma,\tau}(t)$ is the slowly varying envelope and $\omega_{0}=1/\sqrt{L_{\mathrm{g}} C_{\Sigma}}$ is the carrier frequency.
Under this approximation, by retaining only the dominant terms, the second order differential equation is reduced to
\begin{eqnarray}\label{eq:envelope_bulk_A}
\mathrm{i}\frac{\mathrm{d}v_{n}^{\mathrm{L},\mathrm{A}}}{\mathrm{d}t}&=&-\frac{C_{1}+C_{2}+C_{\mathrm{NL}}(v_{n}^{\mathrm{L},\mathrm{A}})}{2C_{\Sigma}}\omega_{0}\,v_{n}^{\mathrm{L},\mathrm{A}}
+\frac{C_{1}}{2C_{\Sigma}}\omega_{0}\,v_{n}^{\mathrm{L},\mathrm{B}}\nonumber\\
&&+\frac{C_{2}}{2C_{\Sigma}}\omega_{0}\,v_{n-1}^{\mathrm{L},\mathrm{B}},
\end{eqnarray}
We further define $V_{n}^{\mathrm{L},\mathrm{A}}(t)=v_{n}^{\mathrm{L},\mathrm{A}}(t)e^{-\mathrm{i}\omega_{0}t}$. The above equation then becomes
\begin{eqnarray}
\mathrm{i}\frac{\mathrm{d}V_{n}^{\mathrm{L},\mathrm{A}}}{\mathrm{d}t}&=&\left[1-\frac{C_{1}+C_{2}+C_{\mathrm{NL}}(V_{n}^{\mathrm{L},\mathrm{A}})}{2C_{\Sigma}}\right]\omega_{0}V_{n}^{\mathrm{L},\mathrm{A}}\nonumber\\
&&+\frac{C_{1}}{2C_{\Sigma}}\omega_{0}V_{n}^{\mathrm{L},\mathrm{B}}+\frac{C_{2}}{2C_{\Sigma}}\omega_{0}V_{n-1}^{\mathrm{L},\mathrm{B}},
\label{eq_app_A_2}
\end{eqnarray}
By following the same approach, the equations for the other circuit nodes can be obtained \cite{PRB113-214310}.

To derive Eqs. (\ref{eq1})-(\ref{eq2}) in the main text, we set $C_{\mathrm{g}}^{\prime}=C_{\mathrm{g}} + C_{2} - C_{3}$ and 
introduce the following parameters: $E_{0}=\left(1-\frac{C_{1}+C_{2}}{2C_{\Sigma}}\right)\omega_{0}$,
$J=\frac{C_{1}}{2C_{\Sigma}}\omega_{0}$, $J^{\prime}=\frac{C_{2}}{2C_{\Sigma}}\omega_{0}$, $J_{d}=\frac{C_{3}}{2C_{\Sigma}}\omega_{0}$,
and $g \left( V_{n}^{\sigma,\tau} \right)=-\frac{C_{\mathrm{NL}} \left( V_{n}^{\sigma,\tau} \right)} {2C_{\Sigma}}\omega_{0}$. With these definitions, the circuit equations can be
expressed in a compact form as shown in Eqs. (\ref{eq1})-(\ref{eq2}).

When investigating the temporal evolutions of voltage intensities under different excitation voltages, we utilize the standard fourth-order Runge-Kutta algorithm 
to directly solve Eqs. (\ref{eq1})-(\ref{eq2}) presented in the main text. A sufficiently small time step is employed to ensure the accuracy of the results.

\section{Experimental details}
\label{app_exp}

To minimize the parasitic effects and tolerance of the circuit components, we utilize capacitors with low equivalent series inductance (ESL) and a tolerance of $\pm 1\%$. We employ inductors featuring magnetic shielding and low series resistance, carefully selecting components with the assistance of an inductance-capacitance-resistance (LCR) meter (HIOKI IM3536). After this selection process, we maintain the tolerance for the inductance at $\pm 1\%$, with the series resistance of the inductors approximately $R_{L_{\mathrm{g}}}=0.6~\Omega$ at $500~\mathrm{kHz}$, which closely matches the eigenfrequencies of the coupled topological edge states. We utilize commercial common-cathode varactor diodes (V60DM45C) and measure the capacitance-voltage relationship when the diodes are excited by an AC voltage source at a frequency of $500~\mathrm{kHz}$ \cite{PRB113-214310}. By fitting the experimental results to the theoretical model $C_\mathrm{v}(v)=C_\mathrm{L}/(1+\left \vert v/v_0 \right\vert)^M$, we determine the values of the parameters $v_{0}$ and $M$.

Although in Sec. \ref{sec_temporal} of the main text, we set the number of unit cells in each SSH circuit chain to $N=120$, we opt for $N=12$ when fabricating the circuit sample, taking into account the presence of experimental circuit dissipation. The circuit lattice is fabricated using standard PCB techniques, ensuring that the inductors are sufficiently spaced to prevent mutual coupling. The PCB traces are designed with a relatively large width of $0.75~\mathrm{mm}$ to accommodate high currents, and the layouts are meticulously optimized to minimize parasitic parameters and coupling with other components.

In the experimental implementation of voltage excitation and temporal evolution, we employ an SPDT switch with two independent channels (ADG1636). This component type is chosen for its low on-resistance of approximately $1~\Omega$, which leads to $R_{L_{\mathrm{g}}}^{\prime}=1.6~\Omega$. To ensure that the left and right SSH circuit chains exhibit a mirror-symmetric dissipation background with respect to the interchain coupling capacitor $C_{3}$, specifically to ensure that the rightmost node of the left chain and the leftmost node of the right chain have equal series resistance, one channel of the SPDT switch, as shown in Fig. \ref{fig3}, is connected to the rightmost node of the left SSH circuit chain, while the other channel is connected to the leftmost node of the right SSH circuit chain. To align with this circuit configuration, the term $-\mathrm{i}\gamma^{\prime} V_{n}^{\sigma,\tau}$ should be included in the equations for both the rightmost site of the left SSH circuit chain and the leftmost site of the right chain in the calculation of theoretical results shown in Fig. \ref{fig4}(b).

The SubMiniature version A (SMA) connectors are connected to external voltage sources. The SPDT switch is powered by $\pm 9~\mathrm{V}$ dual supplies, and the switching sequence is controlled by a digital signal generated from an arbitrary wave function generator (AFG31022). It is important to note that the switch channel connected to the leftmost node of the right SSH circuit chain is configured to ensure that the capacitor $C_{\mathrm{g}}^{\prime}$ and varactor diode $C_{\mathrm{v}}$ are always connected to the inductor $L_{\mathrm{g}}$.

The temporal evolution of the voltage intensities is recorded using an oscilloscope (MDO34). To obtain the envelope of the temporal voltages, a Hilbert transform is applied to the raw data. Additionally, to suppress the effect of circuit dissipation and to facilitate better comparison with the theoretical results, the temporal voltages are normalized with respect to the instantaneous total voltage intensity of the entire circuit lattice. Specifically, the new voltage values are defined as $V_{n}^{\sigma,\tau}(t)/\sqrt{\sum_{n,\sigma,\tau}|V_{n}^{\sigma,\tau}(t)|^{2}}$. Furthermore, we would like to note that the time $t = 35~\mu\mathrm{s}$ selected in Fig. \ref{fig4}(c) is chosen well before the voltage signals become indistinguishable from the background noise.

Finally, it is important to note that the rapid temporal oscillations of the voltage distributions, as shown in Fig. \ref{fig4}(a), are attributed to experimental voltage noise. There are two main sources of this experimental noise. First, the external voltage sources can introduce power-supply ripples. Second, background noise is present in the oscilloscope measurements. This noise becomes apparent when the excitation voltage is low and becomes negligible when the excitation voltage is high.

\section{{Physical mechanism of nonlinearity-controlled temporal routing}}
\label{app_mechanism}

{In this section, to provide deeper insight into the physical mechanism of nonlinearity-controlled temporal routing, we evaluate hybridization with the bulk states of the entire lattice during the routing dynamics and present an effective description of these dynamics based on the two coupled topological edge states.}

\begin{figure}
\centering
\includegraphics[width=9cm]{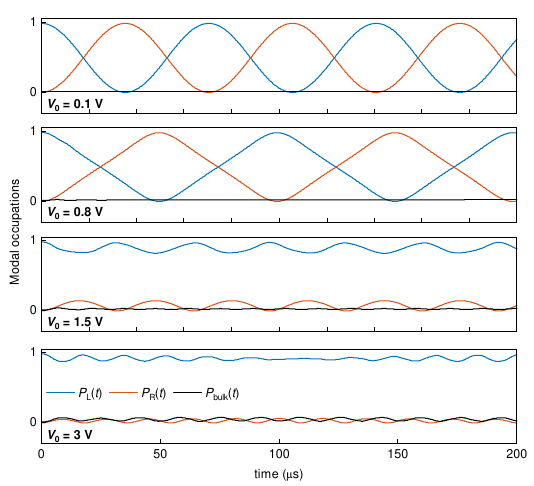}
\caption{{Temporal evolution of the modal occupations in the absence of circuit dissipation.
The rightmost site of the left SSH chain is initially excited with $V_0=0.1$, $0.8$, $1.5$, and $3~\mathrm{V}$, from top to bottom.
The blue and red curves represent the occupations $P_{\mathrm{L}}(t)$ and $P_{\mathrm{R}}(t)$ of the left- and right-localized states $\boldsymbol{\phi}_{\mathrm{L}}$ and $\boldsymbol{\phi}_{\mathrm{R}}$, respectively.
The black curves represent the total bulk-mode occupation $P_{\mathrm{bulk}}(t)$.}}
\label{app_fig_occupation}
\end{figure}

{
We first quantify hybridization with the bulk states during the routing dynamics by projecting the voltage vector $\mathbf{V}(t)$ onto the orthonormal linear eigenmodes $\boldsymbol{\phi}_{\mu}$ of the entire lattice. These eigenmodes are calculated by setting $C_{\mathrm{NL}}=0$ while retaining the interchain coupling $J_d$. The total bulk-mode occupation is defined as~\cite{PRA104-053531}
\begin{equation}
P_{\mathrm{bulk}}(t)=
\frac{\displaystyle\sum_{\mu\in\mathcal{B}}
\left|\boldsymbol{\phi}_{\mu}^{\dagger}\mathbf{V}(t)\right|^{2}}
{\mathbf{V}^{\dagger}(t)\mathbf{V}(t)},
\label{eq_bulk_occupation}
\end{equation}
where $\mathcal{B}$ denotes the set of bulk modes shown as blue dots in Fig.~\ref{fig2}(a), excluding the two coupled topological edge states and the two edge states localized at the outer boundaries. For lossless evolution over the interval $0\leq t\leq200~\mu\mathrm{s}$, according to the results shown in Fig.~\ref{app_fig_occupation}, the maximum bulk-mode occupations are approximately $0.02$, $0.03$, $0.04$, and $0.08$ at $V_0=0.1$, $0.8$, $1.5$, and $3~\mathrm{V}$, respectively. These results indicate that the bulk states make a limited contribution to the routing dynamics within the investigated voltage range.}

{
Next, to examine the contributions of the two coupled topological edge states and explain the routing dynamics based on these states, we construct two orthonormal basis states from the normalized symmetric and antisymmetric states $\boldsymbol{\phi}_{\mathrm{s}}$ and $\boldsymbol{\phi}_{\mathrm{as}}$ of the linear lattice~\cite{PRA104-053531}:
\begin{eqnarray}
\boldsymbol{\phi}_{\mathrm{L}} &=& \frac{\boldsymbol{\phi}_{\mathrm{s}}+\boldsymbol{\phi}_{\mathrm{as}}}{\sqrt{2}},
\label{eq_tm_local_basis_L}\\
\boldsymbol{\phi}_{\mathrm{R}} &=& \frac{\boldsymbol{\phi}_{\mathrm{s}}-\boldsymbol{\phi}_{\mathrm{as}}}{\sqrt{2}}.
\label{eq_tm_local_basis_R}
\end{eqnarray}
The resulting states $\boldsymbol{\phi}_{\mathrm{L}}$ and $\boldsymbol{\phi}_{\mathrm{R}}$ are localized near the right end of the left SSH chain and the left end of the right SSH chain, respectively. Specifically, for the weak interchain coupling considered here, $\boldsymbol{\phi}_{\mathrm{L}}$ and $\boldsymbol{\phi}_{\mathrm{R}}$ closely approximate the topological edge states at the right end of the isolated left chain and the left end of the isolated right chain, respectively. Similarly, the occupations of $\boldsymbol{\phi}_{\mathrm{L}}$ and $\boldsymbol{\phi}_{\mathrm{R}}$ during the temporal evolution are calculated by projecting the voltage vector $\mathbf{V}(t)$ onto these basis states and normalizing by the total voltage intensity:
\begin{eqnarray}
P_{\mathrm{L}}(t) &=&
\frac{\left|\boldsymbol{\phi}_{\mathrm{L}}^{\dagger}\mathbf{V}(t)\right|^{2}}
{\mathbf{V}^{\dagger}(t)\mathbf{V}(t)},
\label{eq_edge_occupation_L}\\
P_{\mathrm{R}}(t) &=&
\frac{\left|\boldsymbol{\phi}_{\mathrm{R}}^{\dagger}\mathbf{V}(t)\right|^{2}}
{\mathbf{V}^{\dagger}(t)\mathbf{V}(t)}.
\label{eq_edge_occupation_R}
\end{eqnarray}}

{
Figure~\ref{app_fig_occupation} shows the temporal evolution of the modal occupations at different excitation voltages.
In Eqs.~(\ref{eq1})-(\ref{eq2}), the voltage-dependent term $g(V)$ modifies the local onsite frequencies.
At $V_0=0.1~\mathrm{V}$, $P_{\mathrm{L}}(t)$ and $P_{\mathrm{R}}(t)$ exhibit nearly complete periodic exchange, with a period approximately equal to $2\pi/\left(\omega_{\mathrm{s}}-\omega_{\mathrm{as}}\right)$. As discussed in the main text, the frequency shifts induced by nonlinearity are weak in this regime, allowing nearly resonant exchange between $\boldsymbol{\phi}_{\mathrm{L}}$ and $\boldsymbol{\phi}_{\mathrm{R}}$.
As $V_0$ increases to $0.8~\mathrm{V}$, the exchange remains nearly complete, but its period increases as the amplitude-dependent nonlinear frequency shifts become significant during the exchange.
At $V_0=1.5$ and $3~\mathrm{V}$, $P_{\mathrm{L}}(t)$ remains larger than $P_{\mathrm{R}}(t)$, indicating that the excitation remains predominantly localized on the left side.
This localization arises because the unequal amplitudes on the two sides produce a substantial frequency mismatch that suppresses resonant exchange.
These results agree with those shown in Fig.~\ref{fig2}(b) and demonstrate a transition from periodic interchain exchange to slowed interchain exchange and ultimately to localization with strongly suppressed interchain exchange.
This behavior is qualitatively consistent with the oscillation and self-trapping dynamics of two weakly coupled Bose--Einstein condensates~\cite{PRA59-620}.}

\section{{Nonlinear lattice composed of two topologically trivial SSH chains}}
\label{app_trivial}

{
To further examine the role of the topological edge states, we interchange the intracell and intercell coupling capacitances by setting $C_1=1500~\mathrm{pF}$ and $C_2=180~\mathrm{pF}$, while retaining $C_3=100~\mathrm{pF}$. All other independent parameters are the same as those used in Sec.~\ref{sec_temporal}, and the condition $C_{\mathrm g}^{\prime}=C_{\mathrm g}+C_2-C_3$ is maintained. Both SSH chains are therefore topologically trivial and do not support topological edge states in the linear limit.}

\begin{figure}
\centering
\includegraphics[width=9cm]{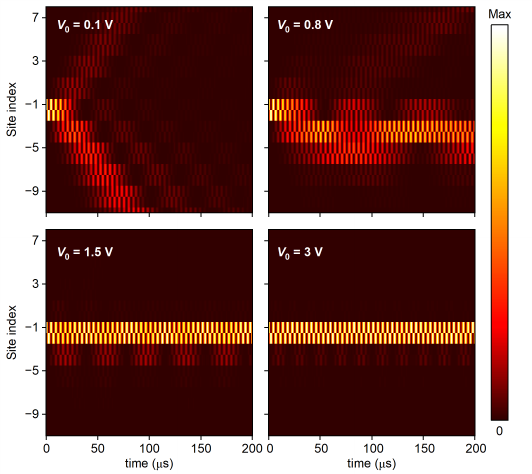}
\caption{{Temporal evolution of voltage intensity distributions when both constituent SSH chains are in the topologically trivial phase.
The intracell and intercell coupling capacitances are set to $C_1=1500~\mathrm{pF}$ and $C_2=180~\mathrm{pF}$, respectively.
The rightmost site of the left circuit chain is initially excited with $V_0=0.1$, $0.8$, $1.5$, and $3~\mathrm{V}$, as indicated in the panels.
The color scale represents the voltage intensity normalized to its maximum value in each panel.
Negative site indices correspond to the left circuit chain, while zero and positive site indices correspond to the right circuit chain.}}
\label{app_trivial_fig}
\end{figure}

{
We numerically solve Eqs.~(\ref{eq1})-(\ref{eq2}) with an initial excitation voltage $V_0$ applied to the rightmost site of the left chain. Figure~\ref{app_trivial_fig} shows the resulting temporal evolution of the voltage intensity distributions. At $V_0=0.1~\mathrm{V}$, the initial excitation spreads into the bulk of both chains, and the localized periodic exchange between the two interface sites observed in Fig.~\ref{fig2}(b) is absent. At $V_0=0.8~\mathrm{V}$, the spatial spreading is reduced, but the excitation still extends into the bulk of the left chain. At $V_0=1.5~\mathrm{V}$ and $3~\mathrm{V}$, the voltage intensity is predominantly confined to the excitation site and its neighboring site in the left chain, with oscillations between these two sites. These sites, indexed as $-1$ and $-2$, belong to the same unit cell and are connected by the strong intracell coupling.}

{
There are clear differences between the dynamics shown in Fig.~\ref{fig2}(b) and Fig.~\ref{app_trivial_fig}. Localized periodic interchain exchange occurs at low excitation voltages when both constituent SSH chains are in the topologically nontrivial phase. In contrast, when both chains are in the topologically trivial phase, such exchange is absent. At higher excitation voltages, both configurations exhibit nonlinear localization, but with different spatial distributions and dynamics. When both chains are topologically trivial, the voltage intensity oscillates between the excitation site and its neighboring site within the left chain, whereas it remains predominantly confined to the excitation site when both chains are topologically nontrivial. These comparisons highlight the role of topological edge states in supporting the temporal routing observed in Fig.~\ref{fig2}(b).}

\section{Model with circuit dissipation}
\label{app_dissipation}

For the fabricated circuit lattice shown in Fig. \ref{fig3}(b), dissipation is unavoidable. First, the inductors $L_{\mathrm{g}}$ have parasitic series resistances $R_{L_{\mathrm{g}}}$, which constitute the primary source of dissipation. Second, the SPDT switch exhibits an on-resistance, which introduces additional dissipation at the excitation node. For simplicity, we combine the on-resistance of the SPDT switch with the series resistances of the inductors at the excitation node, resulting in a new series resistance $R_{L_{\mathrm{g}}}^{\prime}$. 

Following the same approach introduced in Appendix \ref{app_model}, we can formulate the circuit equations for all the nodes in the circuit lattice while including dissipation effects \cite{PRB113-214310}. For instance, for the sublattice site $\mathrm{A}$ in the $n$th unit cell of the left SSH circuit chain with $n\neq N$, we only need to modify Eq. (\ref{eq_app_A}) to incorporate the terms involving $R_{L_{\mathrm{g}}}$. Under the slowly varying envelope approximation, we arrive at the following equation:
\begin{eqnarray}
\mathrm{i}\frac{\mathrm{d}V_{n}^{\mathrm{L},\mathrm{A}}}{\mathrm{d}t}&=&\left[1-\frac{C_{1}+C_{2}+C_{\mathrm{NL}}(V_{n}^{\mathrm{L},\mathrm{A}})}{2C_{\Sigma}}\right]\omega_{0}V_{n}^{\mathrm{L},\mathrm{A}}\nonumber\\
&&-\mathrm{i}\frac{R_{L_{\mathrm{g}}}}{2 L_{\mathrm{g}}} V_{n}^{\mathrm{L},\mathrm{A}} +\frac{C_{1}}{2C_{\Sigma}}\omega_{0}V_{n}^{\mathrm{L},\mathrm{B}} \nonumber \\
&&+\frac{C_{2}}{2C_{\Sigma}}\omega_{0}V_{n-1}^{\mathrm{L},\mathrm{B}}.
\end{eqnarray}
Compared with Eq. (\ref{eq_app_A_2}), the additional term describes the dissipation induced by the series resistance of the inductor. We define $\gamma= \frac{R_{L_{\mathrm{g}}}}{2 L_{\mathrm{g}}}$ and $\gamma^{\prime}= \frac{R_{L_{\mathrm{g}}}^{\prime}}{2 L_{\mathrm{g}}}$. When the circuit equations are expressed in a compact form, we find that, in comparison with Eqs. (\ref{eq1})-(\ref{eq2}), there is an additional term $-\mathrm{i}\gamma^{\prime} V_{n}^{\sigma,\tau}$ in the equation for the excitation site of the circuit lattice, and $-\mathrm{i}\gamma V_{n}^{\sigma,\tau}$ in the equations for the other sites.

Next, we examine the temporal evolution of the voltage distributions when the rightmost site of the left circuit chain is excited with a voltage $V_{0}$. In this case, circuit dissipation is taken into account. We set $R_{L_{\mathrm{g}}}=0.6~\Omega$ and $R_{L_{\mathrm{g}}}^{\prime}=1.6~\Omega$ (the rationale behind these values is explained in Appendix \ref{app_exp}). It is important to note that to conform to the design of the experimentally realized circuit lattice, the term $-\mathrm{i}\gamma^{\prime} V_{n}^{\sigma,\tau}$ is included in the equations for both the rightmost site of the left SSH circuit chain and the leftmost site of the right chain.

When the dissipative terms are included in Eqs. (\ref{eq1})-(\ref{eq2}), the temporal evolution of voltage intensities exhibits damping, resulting in an overall decay of the voltage intensities for any excitation voltage $V_{0}$. To align with the experimental measurement results, and considering that the small voltages obtained experimentally at $t > 40~\mu\mathrm{s}$ for $V_{0} = 0.1~\mathrm{V}$ are indistinguishable from the background noise, we restrict our theoretical voltage distributions to the temporal range of $0 \leq t \leq 40~\mu\mathrm{s}$. Additionally, the theoretical voltage values are also normalized as $V_{n}^{\sigma,\tau}(t)/\sqrt{\sum_{n,\sigma,\tau}|V_{n}^{\sigma,\tau}(t)|^{2}}$. As shown in Fig. \ref{fig4}(b), the transition from periodic interchain exchange to slowed interchain exchange and ultimately to localization with strongly suppressed interchain exchange remains distinctly observable. 
{
These findings demonstrate that the nonlinearity-controlled temporal routing of topological edge states remains observable even when circuit dissipation is taken into account.}

\printcredits

\section*{Data availability}
Data will be made available on request.

\section*{Declaration of competing interest}
The authors declare that they have no known competing financial interests or personal relationships that could have appeared to influence the work reported in this paper.

\section*{Acknowledgment}
L.X., M.I., and R.L. were sponsored by the National Key Research and Development Program of China (Grant No. 2022YFA1404902), the
National Natural Science Foundation of China (Grant No. 12104353), the Natural Science Basic Research Program of Shaanxi (Program No. 2026JC-YBMS-0036), and the Fundamental Research Funds for the Central
Universities (Grant No. QTZX25086). Y.L. was sponsored by the National Natural Science Foundation of China (NSFC) under Grant No. 61871309 and the 111 Project.


\begin{thebibliography}{9}
\bibitem{RMP82-3045} M. Z. Hasan and C. L. Kane, Colloquium: Topological insulators, Rev. Mod. Phys. 82, 3045 (2010).

\bibitem{RMP83-1057} X.-L. Qi and S.-C. Zhang, Topological insulators and superconductors, Rev. Mod. Phys. 83, 1057 (2011).

\bibitem{NRP1-281} G. Ma, M. Xiao, and C. T. Chan, Topological phases in acoustic and mechanical systems, Nat. Rev. Phys. 1, 281 (2019).

\bibitem{NRM7-974} H. Xue, Y. Yang, and B. Zhang, Topological acoustics, Nat. Rev. Mater. 7, 974 (2022).

\bibitem{RPP86-106501} W. Zhu, W. Deng, Y. Liu, J. Lu, H.-X. Wang, Z.-K. Lin, X. Huang, J.-H. Jiang, and Z. Liu, Topological phononic metamaterials, Rep. Prog. Phys. 86, 106501 (2023).

\bibitem{RMP91-015005} N. R. Cooper, J. Dalibard, and I. B. Spielman, Topological bands for ultracold atoms, Rev. Mod. Phys. 91, 015005 (2019).

\bibitem{RMP91-015006} T. Ozawa et al., Topological Photonics, Rev. Mod. Phys. 91, 015006 (2019).

\bibitem{LSA9-1} M. Kim, Z. Jacob, and J. Rho, Recent advances in 2D, 3D and higher-order topological photonics, Light: Science \& Applications 9, 130 (2020).

\bibitem{APR2-2100013} H. Xue, Y. Yang, and B. Zhang, Topological Valley Photonics: Physics and Device Applications, Adv. Photonics Res. 2, 2100013 (2021).

\bibitem{PR1093-1} H. Yang, L. Song, Y. Cao, and P. Yan, Circuit realization of topological physics, Physics Reports 1093, 1 (2024).

\bibitem{APLED} H. Sahin, M. B. A. Jalil, and C. H. Lee, Topolectrical circuits—Recent experimental advances and developments, APL Electronic Devices 1, 021503 (2025).

\bibitem{PRL128-097701} {M. Di Ventra, Y. V. Pershin, and C.-C. Chien, Custodial Chiral Symmetry in a Su--Schrieffer--Heeger Electrical Circuit with Memory, Phys. Rev. Lett. 128, 097701 (2022).}

\bibitem{RPP89-048004} {W. Li, J. Wu, N. Han, R. Zhao, F. Chen, Y. Pan, Y. Ren, H. Chen, Z. Gao, C. Shang, and Y. Yang, Observation of custodial chiral symmetry in memristive topological circuits, Rep. Prog. Phys. 89, 048004 (2026).}

\bibitem{FP20-044204} R. Li, W. Wang, X. Kong, B. Lv, Y. Jia, H. Tao, P. Li, and Y. Liu, Realization of a non-Hermitian Haldane model in circuits, Front. Phys. 20, 044204 (2025).

\bibitem{NRP4-184} O. Breunig and Y. Ando, Opportunities in topological insulator devices, Nat. Rev. Phys. 4, 184 (2022).

\bibitem{NC10-2058} F. Zangeneh-Nejad and R. Fleury, Topological analog signal processing, Nat. Commun. 10, 2058 (2019).

\bibitem{AM30-1803229} Z. Zhang, Y. Tian, Y. Wang, S. Gao, Y. Cheng, X. Liu, and J. Christensen, Directional Acoustic Antennas Based on Valley‐Hall Topological Insulators, Advanced Materials 30, 1803229 (2018).

\bibitem{sa9-eadi} R. Jia, S. Kumar, T. C. Tan, A. Kumar, Y. J. Tan, M. Gupta, P. Szriftgiser, A. Alphones, G. Ducournau, and R. Singh, Valley-conserved topological integrated antenna for 100-Gbps THz 6G wireless, Sci. Adv. 9, eadi8500 (2023).

\bibitem{LPR19-2400567} Z. Zhang, D. Liao, Y. Li, X. Xie, Y. Yang, K. Wang, Z. Dong, E. Li, H. Chen, and F. Gao, Energy-Efficient Integrated Photonic Topological Devices, Laser \& Photonics Reviews 19, 2400567 (2025).

\bibitem{NE5-300} A. Nagulu, X. Ni, A. Kord, M. Tymchenko, S. Garikapati, A. Alù, and H. Krishnaswamy, Chip-scale Floquet topological insulators for 5G wireless systems, Nat. Electron. 5, 300 (2022).

\bibitem{SA10-eadp} W. Deng, W. Zhu, T. Chen, H. Sun, and X. Zhang, Ultrasensitive integrated circuit sensors based on high-order non-Hermitian topological physics, Sci. Adv. 10, eadp6905 (2024).

\bibitem{science358-636} B. Bahari, A. Ndao, F. Vallini, A. El Amili, Y. Fainman, and B. Kanté, Nonreciprocal lasing in topological cavities of arbitrary geometries, Science 358, 636 (2017).

\bibitem{science359-1231} M. A. Bandres, S. Wittek, G. Harari, M. Parto, J. Ren, M. Segev, D. N. Christodoulides, and M. Khajavikhan, Topological insulator laser: Experiments, Science 359, eaar4005 (2018).

\bibitem{nature578-246} Y. Zeng et al., Electrically pumped topological laser with valley edge modes, Nature 578, 246 (2020).

\bibitem{nphys20-852} C. R. Leefmans, M. Parto, J. Williams, G. H. Y. Li, A. Dutt, F. Nori, and A. Marandi, Topological temporally mode-locked laser, Nat. Phys. 20, 852 (2024).

\bibitem{PQW1-8} R. Li and W. Zhang, Topological optical frequency combs, Photonic and Quantum Waves 1, 8 (2026).

\bibitem{PRL109-106402} Y. E. Kraus, Y. Lahini, Z. Ringel, M. Verbin, and O. Zilberberg, Topological States and Adiabatic Pumping in Quasicrystals, Phys. Rev. Lett. 109, 106402 (2012).

\bibitem{LPR10-995} Y. Ke, X. Qin, F. Mei, H. Zhong, Y. S. Kivshar, and C. Lee, Topological phase transitions and Thouless pumping of light in photonic waveguide arrays, Laser \& Photonics Reviews 10, 995 (2016).

\bibitem{nature553-59} O. Zilberberg, S. Huang, J. Guglielmon, M. Wang, K. P. Chen, Y. E. Kraus, and M. C. Rechtsman, Photonic topological boundary pumping as a probe of 4D quantum Hall physics, Nature 553, 59 (2018).

\bibitem{PNAS121} K. Yang, Q. Fu, H. C. Prates, P. Wang, Y. V. Kartashov, V. V. Konotop, and F. Ye, Observation of Thouless pumping of light in quasiperiodic photonic crystals, Proc. Natl. Acad. Sci. U.S.A. 121, e2411793121 (2024).

\bibitem{PRA99-053814} C. Li, W. Zhang, Y. V. Kartashov, D. V. Skryabin, and F. Ye, Bloch oscillations of topological edge modes, Phys. Rev. A 99, 053814 (2019).

\bibitem{LPR12-1700348} Y. Zhang, Y. V. Kartashov, Y. Zhang, L. Torner, and D. V. Skryabin, Resonant Edge‐State Switching in Polariton Topological Insulators, Laser \& Photonics Reviews 12, 1700348 (2018).

\bibitem{OL44-3342} H. Zhong, Y. V. Kartashov, Y. Zhang, D. Song, Y. Zhang, F. Li, and Z. Chen, Rabi-like oscillation of photonic topological valley Hall edge states, Opt. Lett. 44, 3342 (2019).

\bibitem{PRA104-053531} N. K. Efremidis, Topological photonic Su-Schrieffer-Heeger-type coupler, Phys. Rev. A 104, 053531 (2021).

\bibitem{LPR14-1900193} W. Song, W. Sun, C. Chen, Q. Song, S. Xiao, S. Zhu, and T. Li, Robust and Broadband Optical Coupling by Topological Waveguide Arrays, Laser \& Photonics Reviews 14, 1900193 (2020).

\bibitem{OL49-6377} {Q. Zhang, J. Kang, R. Wei, and G. Dong, All-optical switching in nonlinear topological waveguide arrays, Opt. Lett. 49, 6377 (2024).}

\bibitem{OL46-6067} N. Bazhan, B. Malomed, and A. Yakimenko, Josephson oscillations of edge quasi-solitons in a photonic-topological coupler, Opt. Lett. 46, 6067 (2021).

\bibitem{nano11-3653} A. A. Arkhipova et al., Observation of nonlinearity-controlled switching of topological edge states, Nanophotonics 11, 3653 (2022).

\bibitem{LPR17-2300024} {S. K. Ivanov, S. A. Zhuravitskii, N. N. Skryabin, I. V. Dyakonov, A. A. Kalinkin, S. P. Kulik, Y. V. Kartashov, V. V. Konotop, and V. N. Zadkov, Macroscopic Zeno Effect in a Su--Schrieffer--Heeger Photonic Topological Insulator, Laser \& Photonics Reviews 17, 2300024 (2023).}

\bibitem{LSA13-99} Z. Zhang, Y. Li, X. Sun, and X. Shu, Visual observation of photonic Floquet–Bloch oscillations, Light: Science \& Applications 13, 99 (2024).

\bibitem{PRL42-1698} W. P. Su, J. R. Schrieffer, and A. J. Heeger, Solitons in Polyacetylene, Phys. Rev. Lett. 42, 1698 (1979).

\bibitem{OL51-1649} R. Li, L. Xu, M. Imran, W. Wang, Y. Jia, and Y. Liu, Symmetry-breaking bifurcation of coupled topological edge states, Opt. Lett. 51, 1649 (2026).

\bibitem{PRR8-L012} K. Sone and Y. Hatsugai, Topological-to-topological transition induced by on-site nonlinearity in a one-dimensional topological insulator, Phys. Rev. Research 8, L012045 (2026).

\bibitem{PRA113-053506} R. Li, M. Imran, W. Wang, Y. Jia, and Y. Liu, Dark solitons in nonlinear Su-Schrieffer-Heeger lattices, Phys. Rev. A 113, 053506 (2026).

\bibitem{book} J. K. Asb\'{o}th, L. Oroszl\'{a}ny, and A. P\'{a}lyi, A Short Course on Topological Insulators, Lecture Notes in Physics, Vol. 919 (Springer International Publishing, Cham, 2016).

\bibitem{CP8} R. Li, X. Kong, W. Wang, Y. Wang, Y. Jia, H. Tao, P. Li, Y. Liu, and B. A. Malomed, Observation of edge solitons and transitions between them in a trimer circuit lattice, Commun. Phys. 8, 342 (2025).

\bibitem{CSF207} R. Li, W. Wang, Y. Jia, Y. Liu, P. Li, and B. A. Malomed, Nonlinear quadrupole topological insulators, Chaos, Solitons \& Fractals 207, 118044 (2026).

\bibitem{PRB113-214310} R. Li, W. Wang, X. Kong, C. Shang, Y. Jia, G.-G. Liu, H. Tao, Y. Liu, and B. Zhang, Nonlinear topological edge states, topological gap solitons, and self-induced topological edge states in nonlinear Su-Schrieffer-Heeger circuit lattices, Phys. Rev. B 113, 214310 (2026).



\bibitem{PRA59-620} S. Raghavan, A. Smerzi, S. Fantoni, and S. R. Shenoy, Coherent oscillations between two weakly coupled Bose--Einstein condensates: Josephson effects, $\pi$ oscillations, and macroscopic quantum self-trapping, Phys. Rev. A 59, 620 (1999).

\end{thebibliography}
\end{document}